\documentclass[prx,twocolumn,superscriptaddress,showpacs,amsmath,amstex,amssymb,citeautoscript]{revtex4-2}
\pdfoutput=1 

\usepackage{dcolumn}
\usepackage{bm}
\usepackage[T1]{fontenc}
\usepackage{braket}
\usepackage{graphicx}
\usepackage{xcolor}
\usepackage{lipsum}
\usepackage[normalem]{ulem}
\usepackage{amsmath}
\usepackage{amsfonts}
\usepackage{amssymb} 
\usepackage{color}
\usepackage[percent]{overpic}
\usepackage{soul} 
\usepackage{amssymb}
\usepackage{wasysym}
\usepackage{dsfont}
\usepackage{float}
\usepackage[mathscr]{euscript}
\usepackage{ifthen}
\usepackage{csquotes}
\usepackage{appendix}
\usepackage[bottom]{footmisc}
\usepackage{verbatim}
\usepackage{multirow}
\usepackage{mathrsfs}
\usepackage[scr=esstix,cal=boondox]{mathalfa} 
\usepackage{CJKutf8}

\renewcommand{\vec}[1]{\mathbf{{#1}}}

\begin{document}

\title{Spin dynamics dominated by superexchange via virtual molecules}


\author{Yoo Kyung Lee}
\altaffiliation{These authors contributed equally to this work.}
\affiliation{Department of Physics, Massachusetts Institute of Technology, Cambridge, MA 02139, USA}
\affiliation{Research Laboratory of Electronics, Massachusetts Institute of Technology, Cambridge, MA 02139, USA}
\affiliation{MIT-Harvard Center for Ultracold Atoms, Cambridge, MA, USA}

\author{Hanzhen Lin (\begin{CJK*}{UTF8}{gbsn}林翰桢\end{CJK*})}
\altaffiliation{These authors contributed equally to this work.}
\affiliation{Department of Physics, Massachusetts Institute of Technology, Cambridge, MA 02139, USA}
\affiliation{Research Laboratory of Electronics, Massachusetts Institute of Technology, Cambridge, MA 02139, USA}
\affiliation{MIT-Harvard Center for Ultracold Atoms, Cambridge, MA, USA}

\author{Wolfgang Ketterle}
\affiliation{Department of Physics, Massachusetts Institute of Technology, Cambridge, MA 02139, USA}
\affiliation{Research Laboratory of Electronics, Massachusetts Institute of Technology, Cambridge, MA 02139, USA}
\affiliation{MIT-Harvard Center for Ultracold Atoms, Cambridge, MA, USA}

\begin{abstract}
\noindent
Simple, paradigmatic systems are important tools in understanding strongly correlated systems. One such system is the Bose-Hubbard model, which can be realized using atoms in optical lattices with delta-function interactions. We report the first experimental observation of two features of the Bose-Hubbard model: superexchange via virtual molecules in excited bands and off-site contact interactions. Adiabatic sweeps confirm the creation of such molecules, exemplifying a new kind of Feshbach resonance. 
\end{abstract}

\maketitle

\noindent {\bf Introduction}.
Important properties of many-body physics can be illuminated by studying paradigmatic models. One important model for materials are particles with short-range interactions in a periodic potential, captured by the Hamiltonian
\begin{equation}
  H = \sum_i \left[ \frac{p_i^2}{2m} + V_z \sin^2(kx_i) \right] +  \sum_{i\neq j}g_{ij}\delta(x_i-x_j)
    \label{eq:sin_potential}
\end{equation}
where $i$ refers to the $i$th particle in a lattice of depth $V_z$ and interaction strength $g_{ij}$ between particles $i,\,j$. This model can be almost perfectly realized in experiments with ultracold atoms in optical lattices, which therefore serve as well-controlled quantum simulators for many-body physics \cite{Jaksch1998,Bloch2012,Georgescu2014,Gross2017,Schafer2020}. The simplest formulation of this model, the standard Hubbard model, assumes single band, nearest-neighbour tunnelling and on-site interactions. However, it is important to study higher-order interaction processes in this model because not only do they lead to corrections for quantitative quantum simulations, but they can also give rise to new phenomena and material properties. Several extensions of the basic Bose-Hubbard model \cite{Fisher1989,Greiner2003} have been featured in previous work, including density-dependent tunnelling \cite{Mazzarella2006,Luhmann2012,Jurgensen2014}, higher-band admixtures via on-site interactions \cite{Campbell2006}, and next-neighbour interactions by dipolar forces \cite{Yan2013,Newman2018}. 

We report two new interaction processes for the basic Hamiltonian Eq.~\eqref{eq:sin_potential}: off-site interactions and nearest-neighbour interactions via virtual molecules. Both exemplify the new physics which emerge when the short-range interactions in Eq.~\eqref{eq:sin_potential} become strong near Feshbach resonances. In the limit of weak lattice potentials, approaching a Feshbach resonance creates complex phenomena where multiple bands are occupied and leads ultimately to the unitarity-limited  Bose gas which has an extremely short lifetime due to recombination processes \cite{Makotyn2014,Fletcher2017,Eigen2018}.

In this work, we look at the lattice physics in the limit of a two-component Mott insulator with one atom per site. The dynamics of this system arises purely from the spin degree of freedom and can be described by an anisotropic Heisenberg model \cite{Duan2003,Jepsen2020}. We use spin dynamics to determine the spin-spin interactions when the interaction term in Eq.~\eqref{eq:sin_potential} is tuned across a Feshbach resonance. With the chosen system, we can create arbitrarily strong interactions since three-body losses are negligible in an $N\,{=}\,1$ Mott insulator. 
It is in this unusual regime that we are able to observe the effects of two new processes. One of them is off-site interactions: since the lowest band Wannier function does not fully localize atoms on one site, Eq.~\eqref{eq:sin_potential} leads to nearest-neighbour contact interactions. This is one of the simplest extensions of the Hubbard model which takes into account effects of longer range interactions, and can cause charge or spin density waves \cite{Clay1999}. More dramatically, we observe strong modifications of spin dynamics for special values of the scattering length. On first sight, it is surprising that many-body dynamics is affected by scattering lengths as large as $15,000\,a_0$, which exceeds even the lattice spacing ($\sim10,000\,a_0$) where $a_0$ is the Bohr radius.  We can explain our observations by a process which has not been predicted before: superexchange through excited molecular branches.
\begin{figure*}[t] 
    \vspace{-10pt}
    \includegraphics[width=\linewidth,keepaspectratio]{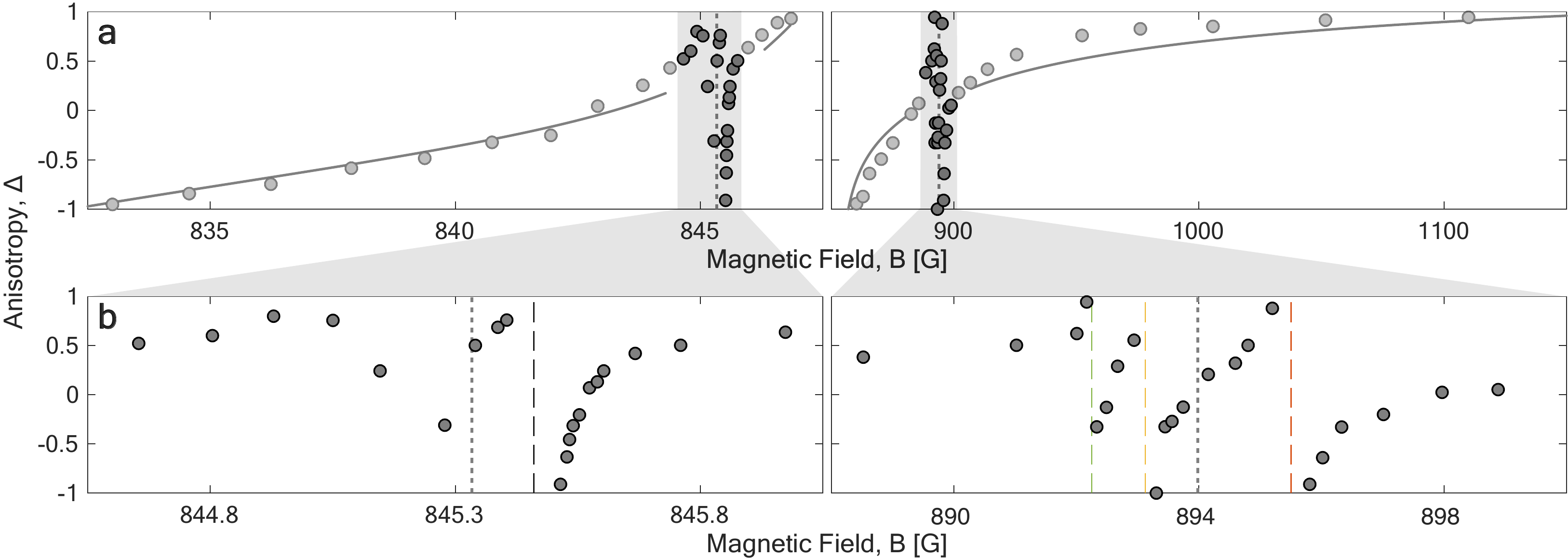}
    \caption{
        {\bf Spin interaction anisotropy at different magnetic fields}. Anisotropy $\Delta=J_z/J_{xy}$ as a function of magnetic field; plotting all data ({\bf a}) and zooming in for fields with large ${bb}$ scattering lengths $|a_{bb}|>500\,a_0$ ({\bf b}). Far from the ${bb}$ Feshbach resonances (vertical dotted lines), the data (grey points) agree well with calculated values (solid grey lines). Data where linear perturbation theory breaks down (black points) shows resonant behavior. We see multiple dispersive features at $845.460\,\text{G},\ 892.250\,\text{G},\ 893.123\,\text{G}$, and $895.504\,\text{G}$ (dashed vertical lines). These divergences can only be qualitatively explained by virtual molecule-mediated superexchange. We use the fits from \cite{Secker2020} with a small correction for the position of the narrow Feshbach resonance by $170\,{\rm mG}$ based on RF spectroscopy data taken close to the resonance (to be published). Note that the behavior of the anisotropy is smooth across the Feshbach resonance. The statistical errors of the data are smaller than the symbols.
        }
	\vspace{-10pt}
	\label{fig:DeltaVsField}
\end{figure*}

Superexchange is a second-order tunnelling process involving intermediate states which are typically virtual doublons in the lowest Bloch band. However, superexchange can also occur through an {\it infinite} number of intermediate states which are not necessarily in the lowest band. One important intermediate state that has been neglected so far in the literature is that of molecules with excitations in their center-of-mass (COM). Due to their large binding energies, such processes should be suppressed; however, when the negative binding energy of a molecule is compensated for by positive kinetic energy of its COM, the energy defect can be small enough that this process becomes greatly enhanced and even resonant. Indeed, by carefully controlling the binding energy using a Feshbach resonance, we observe that spin dynamics can be completely dominated by the temporary formation of virtual molecules in excited bands of the lattice, which have a total energy close to zero and are therefore energetically accessible to the Mott insulator ground state. It is remarkable that the first evidence for such states has been found here through many-body spin dynamics, and not in any of the previous spectroscopic studies of two interacting atoms in trapping potentials \cite{Kohl2006,Ospelkaus2006,Amato-Grill2019,Secker2020}.\\

\noindent {\bf Single-band Bose-Hubbard model}.
The single-band Bose-Hubbard Hamiltonian with two components ${a}$ and ${b}$ including only nearest-neighbour tunnelling is given by
\begin{equation}
\begin{aligned}
    H
    & = - \sum_{i,\sigma} \Big[\,t\,(c_{i,\sigma}^\dagger c_{i+1,\sigma} + {\rm h.c.}) \\
    & + \frac{U_{\sigma\sigma}}{2}n_{i,\sigma}(n_{i,\sigma}-1)
    + V_{\sigma\sigma}n_{i,\sigma}n_{i+1,\sigma} \Big] \\
    & + \sum_i \Big[ U_{ab} n_{i,{a}}n_{i,{b}} + V_{ab} ( n_{i,{a}}n_{i+1,{b}} + {a} \leftrightarrow {b}) \Big]
\end{aligned}
\label{eq:boseHubbard}
\end{equation}
where $c_{i,\sigma} \ (c_{i,\sigma}^\dagger)$ are the annihilation (creation) operators on lattice site $i$ of species $\sigma \in \{ {a}, {b} \}$, and $n_{i,\sigma}$ are the number operators on site $i$. Eq.~\eqref{eq:boseHubbard} is fully parameterized by the tunnelling matrix element $t$, the on-site interaction energies $U_{\sigma\sigma'}$ for $\sigma,\,\sigma' \in \{{a}, {b} \}$, and off-site interactions $V_{\sigma\sigma'}$. Because mass transport is frozen out in a Mott insulator, the only degree of freedom that remains is spin, i.e.~the $S_z$-magnetization of a particle, which must be conserved. We can thus map the two-component Bose-Hubbard Hamiltonian Eq.~\eqref{eq:boseHubbard} onto the anisotropic Heisenberg Hamiltonian by defining ${a}$ as spin $\ket{\downarrow}$ and ${b}$ as spin $\ket{\uparrow}$ \cite{Duan2003,Jepsen2020}, yielding
\begin{equation}
    H_{\rm Heis} 
    = \sum_i [ J_{xy}  (S^x_i S^x_{i+1} + S^y_iS^y_{i+1}) + J_z( S^z_iS^z_{i+1})] 
    \label{eq:heisenberg}
\end{equation}
The anisotropy $\Delta:=J_z/J_{xy}$ is comprised of superexchange and off-site energies $J$ and $V$, respectively \cite{Kuklov2003,Jepsen2020}
\begin{equation}
    \label{eq:JzJxy}
    \begin{aligned}
        J_z \ &= (J_{ab} - 4 V_{ab}) - (J_{aa} - 2 V_{aa}) - (J_{bb} - 2V_{bb})
        \\
        J_{xy} &= -J_{ab}
    \end{aligned}
\end{equation}
where $J_{\sigma\sigma'} := 4t^2 / U_{\sigma\sigma'}$. Perturbatively, the on-site and off-site energies scale linearly with the coupling strength $g_{\sigma\sigma'} = 4\pi\hbar^2 a_{\sigma\sigma'}/m$, which is is a function of the $s$-wave scattering length $a_{\sigma\sigma'}$ and parameterizes the strength of the interaction between two particles. All scattering lengths are well-known from \cite{Secker2020}. The bare tunnelling $t$ in the lowest band is a constant and approximately $0.0164\,E_R$ for a lattice depth of $V_z = 10.7\,E_R$, where $E_R=h^2/8m\lambda_\text{lat}^2\approx25\,{\rm kHz}$ is the recoil energy, $h$ the Planck constant, $m$ the mass of the particle, and $\lambda_\text{lat}=532\,\rm nm$ the lattice spacing.

The superexchange coupling is usually calculated with the tunnelling constant and lowest-band on-site interactions $J_{\sigma\sigma'}=4t^2/U_{\sigma\sigma'}$ and describes a second-order process via an intermediate state of virtual doublons \cite{Duan2003} (with on-site interaction energy $U_{\sigma\sigma'}$ typically occupying the lowest Bloch band). However, a full description of superexchange requires summing over all possible intermediate states; in the following, we will show that multiple bands are required to explain the dynamics of the system near a Feshbach resonance.\\

\begin{figure}[t]
    \centering
    \includegraphics[width=\linewidth,keepaspectratio]{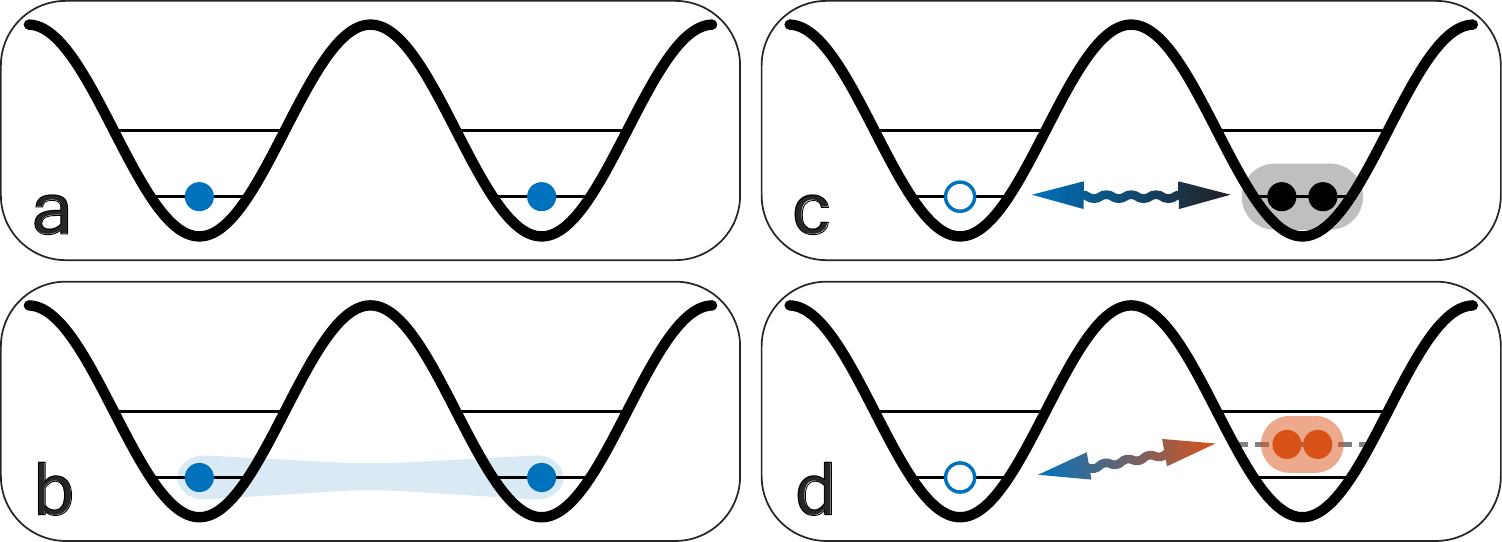}
    \caption{{\bf Different coupling processes between neighbouring lattice sites}. In the singly occupied Mott insulator ({\bf a}) near a Feshbach resonance, the atoms in neighbouring sites can interact directly through the leakage of their Wannier functions (off-site interaction,  ({\bf b}). Alternatively, they can undergo superexchange in which one atom virtually tunnels to the neighbouring site and back. The doubly occupied virtual state is commonly a doublon in the lowest band ({\bf c}) but can also be a loosely bound molecule with excited COM motion ({\bf d}).
    } 
    \label{fig:coupling_processes}
\end{figure}

\begin{figure}[t]
    \includegraphics[width=\linewidth,keepaspectratio]{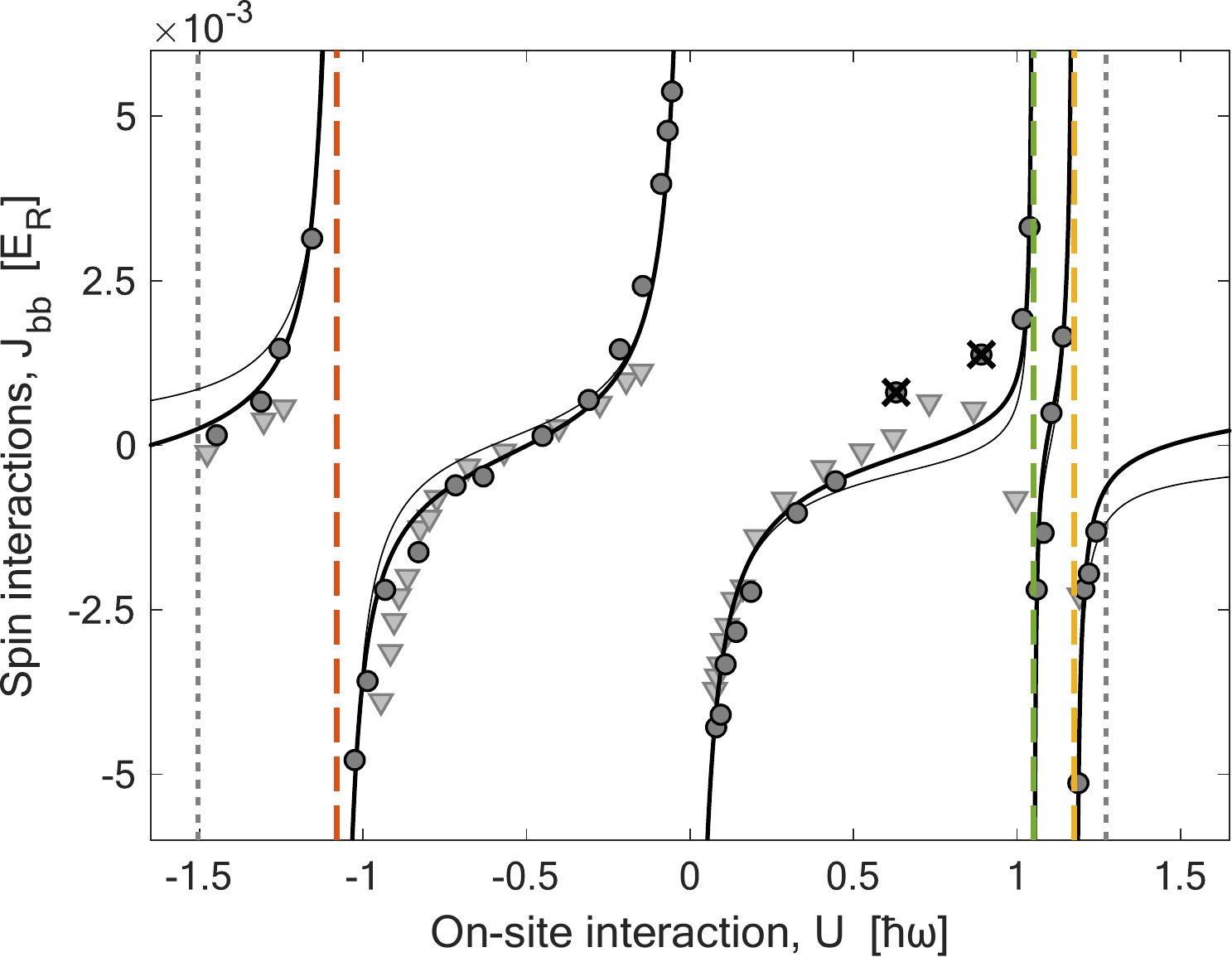}
    \caption{
        {\bf Spin interactions $J_{bb}^*$ as a function of on-site interaction energy of an anisotropic lattice}. When $J_{bb}^*$ is plotted as a function of the lowest-band on-site interaction energy, the data from both broad (circles) and narrow (triangles) resonances collapse onto the same curves. Here, $\hbar\omega\approx4.9\,E_R$ is the gap for a $10.7\,E_R$ lattice. For small $U$, the data shows the $1/U$ scaling of typical superexchange. For larger $U$, three other resonances (dashed vertical lines) appear far from the bulk Feshbach resonances (vertical dotted lines). We plot two fits using Eq.~\eqref{eq:fitFunction} (for the broad resonance only): the sum of four hyperbolas with (solid line, $v\neq0$) and without (thin line, $v=0$) a linear term, which we expect to arise due to the off-site interaction $V_{bb}$. We predict the slope to be $v\approx 1.5$ and we obtain $v=1.3(4)$ ($1\sigma$ value). 
        The two crossed-out points are outliers that were excluded from the fit which might be related to a region of $\Delta$ with higher systematic errors that overestimate the value of $\Delta$ (see \cite{Jepsen2022}).
        }
	\label{fig:Jbb_vs_Ubb}
\end{figure}

\noindent {\bf Experimental methods}.
Our quantum simulator uses ultracold lithium-7 atoms in an optical lattice to realize the two-component Bose-Hubbard model. The atoms are in the lowest and second-lowest hyperfine states, labelled as states ${a}$ and ${b}$ in the following.
This two-level system forms the pseudo-spin and simulate the anisotropic Heisenberg model Eq.~\eqref{eq:heisenberg} when loaded into the $N\,{=}\,1$ Mott insulator phase. The system is turned quasi-1D by lowering the lattice depth along the $z$-direction ($V_{x,y,z} = 35\,E_R,\,35\,E_R,\,10.7\,E_R$). As experimentally demonstrated in previous work \cite{Jepsen2022}, there exist special quasi-stable spin helix states in the 1D anisotropic Heisenberg model whose wavevector $Q_p$ is related to the anisotropy by $\Delta=\cos(Q_p \lambda_{\rm lat})$ \cite{Popkov2021}. By measuring the decay of transverse spin helix states, we can measure the anisotropy $\Delta$ for different magnetic fields as shown in Fig.~\ref{fig:DeltaVsField}. We measure the anisotropy $\Delta$ at magnetic fields $B$ far from the ${aa}$ and ${ab}$ Feshbach resonances and near both the narrow and broad ${bb}$ Feshbach resonances. In this range of fields, $J_{aa}$ and $J_{ab}$ are well-understood and vary only slowly with $B$. We subtract their contributions to $J_z$ and $J_{xy}$ defined by Eq.~\eqref{eq:JzJxy} to isolate the effects of strong interactions between two ${b}$ atoms. This results in a term that depends on only ${bb}$ interactions:
\begin{equation}
    \begin{aligned}
    J_{bb}^* & :=\Delta \cdot J_{xy} + (J_{aa} - 2V_{aa}) - (J_{ab} - 4 V_{ab})
    \\
    & = -J_{bb} + 2 V_{bb}.
    \end{aligned}
    \label{eq:Jbb}
\end{equation}
We analyze primarily the magnetic field in the range  $894\,{\rm G} \pm 5\,{\rm G}$ around the broad ${bb}$ Feshbach resonance. The narrow resonance at $845.3\,\rm G$ has a large effective range, so we expect to see quantitative discrepancies between the two resonances. Henceforth, we drop the subscripts for brevity ($a_{bb}\,{\rightarrow}\,a$, $U_{bb}\,{\rightarrow}\,U$). We measure the anisotropy $\Delta$ at fields extremely close to the resonance, where $|a| > 500\,a_0$ (Fig.~\ref{fig:DeltaVsField}b). Systems in these regions are typically difficult to study due to detrimental three-body loss which scales with $a^4$ \cite{Clay1999,Fedichev1996,Weber2003}. However, the $N\,{=}\,1$ Mott insulator is unique because doubly occupied sites occur only virtually; consequently, triply occupied sites will almost never occur, making our system effectively immune to such losses. This allows us to tune our magnetic field very close to a Feshbach resonance and shed light on how strongly-interacting atoms behave in a lattice. Not only do we observe non-monotonic behavior that cannot be fully explained by off-site interactions, but we also see multiple dispersive features as a function of interaction strength (Figs.~\ref{fig:DeltaVsField}b and \ref{fig:Jbb_vs_Ubb}). We propose that superexchange mediated by virtual molecules cause such dynamics. \\

\begin{figure}[t]
    \centering
    \includegraphics[width=\linewidth]{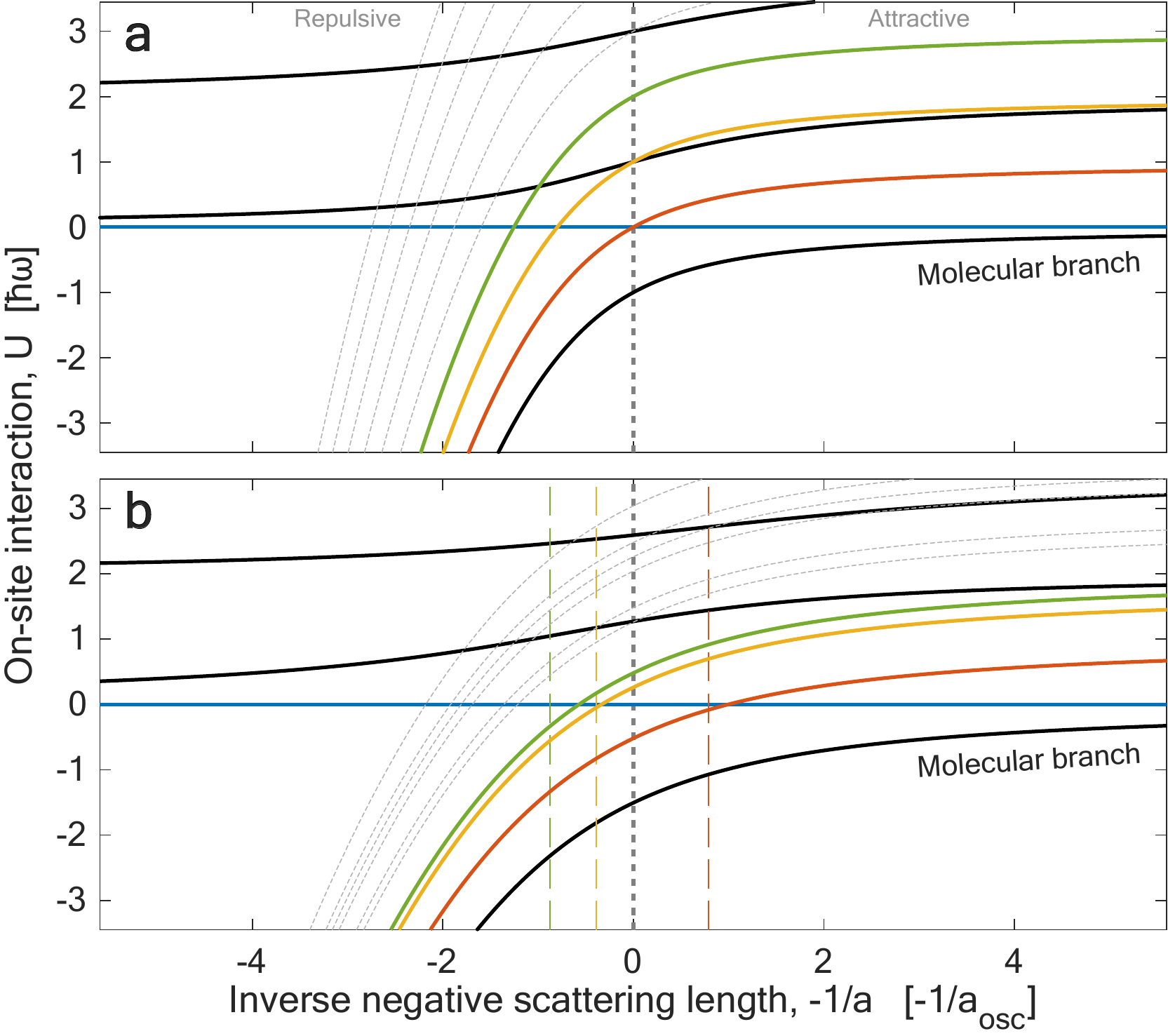}
    \vspace{-15pt}
    \caption{
    {\bf Energy spectrum for two interacting atoms in a trap}.
     Energy spectrum in an isotropic harmonic trap ({\bf a}). Only the relative motion is affected by the scattering length $a$ \cite{Busch1998} (black). Multiple copies of this band structure exist for each unit of COM excitation $\hbar\omega$. We emphasize the lowest three excitations (red, yellow, green) for the lowest molecular branch, which correspond to excitations of $E_{\rm COM}=\hbar\omega,\,2\hbar\omega$, and $3\hbar\omega$, respectively. These branches cross zero energy (i.e.~are degenerate with non-interacting atoms, horizontal blue line) near a Feshbach resonance ($[a/a_{\rm osc}]^{-1}=0$, vertical dotted line). 
     Energy spectrum in an anisotropic harmonic trap ({\bf b}) \cite{Idziaszek2006}, in which the radial frequency is 1.8 times greater than the axial frequency ($\omega_x=\omega_y=1.8\,\omega_z$), consistent with experimental conditions. Like in {\bf a}, the lowest branch of relative motion with COM excitations are plotted (black) as well as the three lowest COM excitations ($E_{\rm COM}=\hbar\omega_z,\,\hbar\omega_x,\,2\hbar\omega_z$). In the anisotropic trap, the first excited branch (red) does not cross zero energy at the Feshbach resonance, but at negative scattering length (strongly attracting doublons). The second and third excited branches (yellow and green) cross zero energy at positive scattering lengths (loosely bound molecules). Vertical dashed lines correspond to the fitted resonance positions in Fig.~\ref{fig:DeltaVsField}b for the broad resonance only; the positions of all three crossings are in qualitative agreement with the predictions of the anisotropic harmonic model, and the first two (red, yellow) agree quantitatively as well. Thin dashed lines correspond to the next six higher COM excitations for both panels.
    }
    \vspace{-5pt}
    \label{fig:trap_branches}
\end{figure}

\noindent {\bf Virtual molecule-mediated superexchange}. The energy spectrum of two identical particles with contact interactions has been solved analytically for isotropic three-dimensional harmonic traps \cite{Busch1998}. An important step in this solution is the separation of the relative motion and the COM motion, as only the relative motion is affected by interactions and is orthogonal to the COM motion. Thus, the combined energy spectrum of both COM and relative motion will lead to an infinite number of up-shifted copies of the relative motion, whose energy shifts correspond to excitations in the COM degree of freedom. Fig.~\ref{fig:trap_branches}a shows the energy spectrum of the relative motion, as well as the relative motion including the lowest nine COM excitations. We note that the first excited molecular branch crosses zero energy at the Feshbach resonance ($[a/a_{\rm osc}]^{-1} = 0$).

A cubic optical lattice is comprised of many traps. Two atoms in completely isolated traps have an interaction energy of zero. If tunnelling is allowed, one atom can tunnel to the site of the other atom; thus, the state of two atoms in one trap is connected to the state of non-interacting isolated atoms via tunnelling. The directionality of tunnelling can yield excitations in the COM motion. In other words, when one atom tunnels into an excited band on the site of a neighbouring atom, it can generate excitations in the COM motion of the pair.

Of significant interest are the implications of the zero-crossings of the excited branches to superexchange. The superexchange $J$ arises from second-order perturbation theory, and should include different possible intermediate states. Nevertheless, most experiments work in the regime of small scattering length and the superexchange process is dominated by the branch with no COM excitation (Fig.~\ref{fig:trap_branches}, solid black lines) as it possesses the smallest energy defect. However, in the limit of large scattering lengths, the positive COM excitation and the negative quasi-molecular binding energy can cancel and the net energy can cross zero, resulting in resonantly enhanced superexchange interactions. These intermediate states are either strongly attracting doublons or loosely bound molecules in excited motional branches, which have been neglected in previous works but becomes the dominant contributor in this regime. In reality, our traps more closely resemble anisotropic harmonic potentials whose energy levels are plotted in Fig.~\ref{fig:trap_branches}b \cite{Idziaszek2006}. The net energy of the branch with one unit of COM energy crosses zero slightly before it reaches the fully molecular regime (Fig.~\ref{fig:trap_branches}b, red), while several other branches cross zero slightly after (Fig.~\ref{fig:trap_branches}b, yellow and green). The positions of the zero crossings using the anisotropic harmonic potentials are in good qualitative agreement with the positions of the resonances that we fit (Fig.~\ref{fig:trap_branches}b, dashed lines). The quantitative discrepancy of the third-lowest resonance position (Fig.~\ref{fig:trap_branches}b, green) could be explained by the anharmonicity of our optical lattice potential which be deconfining compared to a harmonic trap.\\

\noindent {\bf Off-site interactions}.
Thus far, higher-order corrections to the Hubbard Hamiltonian such as bond-charge tunnelling \cite{Luhmann2012,Jurgensen2014} and nearest-neighbour dipolar interactions \cite{Yan2013,Newman2018} have been observed. Nearest-neighbour {\it contact} interactions (Fig.~\ref{fig:coupling_processes}b), which we refer to as off-site interactions $V$, have not been observed until now. Indeed, because their strength is determined by the overlap of two {\it neighbouring} Wannier functions, off-site interactions are much weaker than on-site interactions (which are determined by the overlap of two Wannier functions on the {\it same} site); we define the ratio of overlaps $\kappa\approx 6.915\times10^{-5}$. Thus, observing these interactions was thought to be impossible. In a Mott insulator, however, only spin interactions can occur; therefore, the energy scale of interest is {\it not} given by the on-site interaction energies $U$ but by the superexchange energy scale $J \propto t^2/U$. Therefore, near a Feshbach resonance where interactions grow large, regular superexchange processes will be suppressed while off-site interactions will be enhanced, which helps us to observe the latter.\\

\noindent
{\bf Fitting.}
Fig.~\ref{fig:Jbb_vs_Ubb} connects the observed spin phenomena to our existing knowledge of Bose-Hubbard interactions. We map $J_{bb}^*$, the isolated contribution to the spin interaction from two $b$ atoms, to the on-site energy $U$ expected from a lowest-band description. In this parameterization, superexchange terms will be hyperbolic, whereas the off-site interaction---a weakened version of the on-site interaction---are expected to be linear . Superexchange via virtual molecules in excited bands will cause divergences at $U\neq 0$ as the resonance condition is shifted by COM excitations. The measured data is well explained by the sum of four hyperbolas and a linear term:
\begin{equation}
    J_{bb}^*(U)
    \approx \frac{-4 t^2}{U} + \sum_{n = 1}^3 \frac{-A_n}{U-U_n} + B \cdot \kappa\,U.
\label{eq:fitFunction}
\end{equation}
$U_n$, $t$, $A_n$, and $B$ are treated as free fit parameters. The first term describes the usual superexchange process. We propose that the off-site interactions scale linearly with on-site interactions, reduced by a factor $\kappa$ that accounts for reduced wavefunction overlap. This treatment predicts $B\approx 1.5$, and our data fits $B=1.3(4)$ ($1\sigma$ value); the fitted value is in good agreement with our prediction and is more than three times the standard deviation, which suggests our off-site interaction is statistically significant. However, the signature of off-site interactions is small compared to the new resonance features (Fig.~\ref{fig:Jbb_vs_Ubb}); the fact that we can observe this weak feature at all speaks to the high quality of our measurements of the anisotropy $\Delta$ \cite{Jepsen2022}.\\

\begin{figure}[t]
    \centering
    \includegraphics[width=1\linewidth]{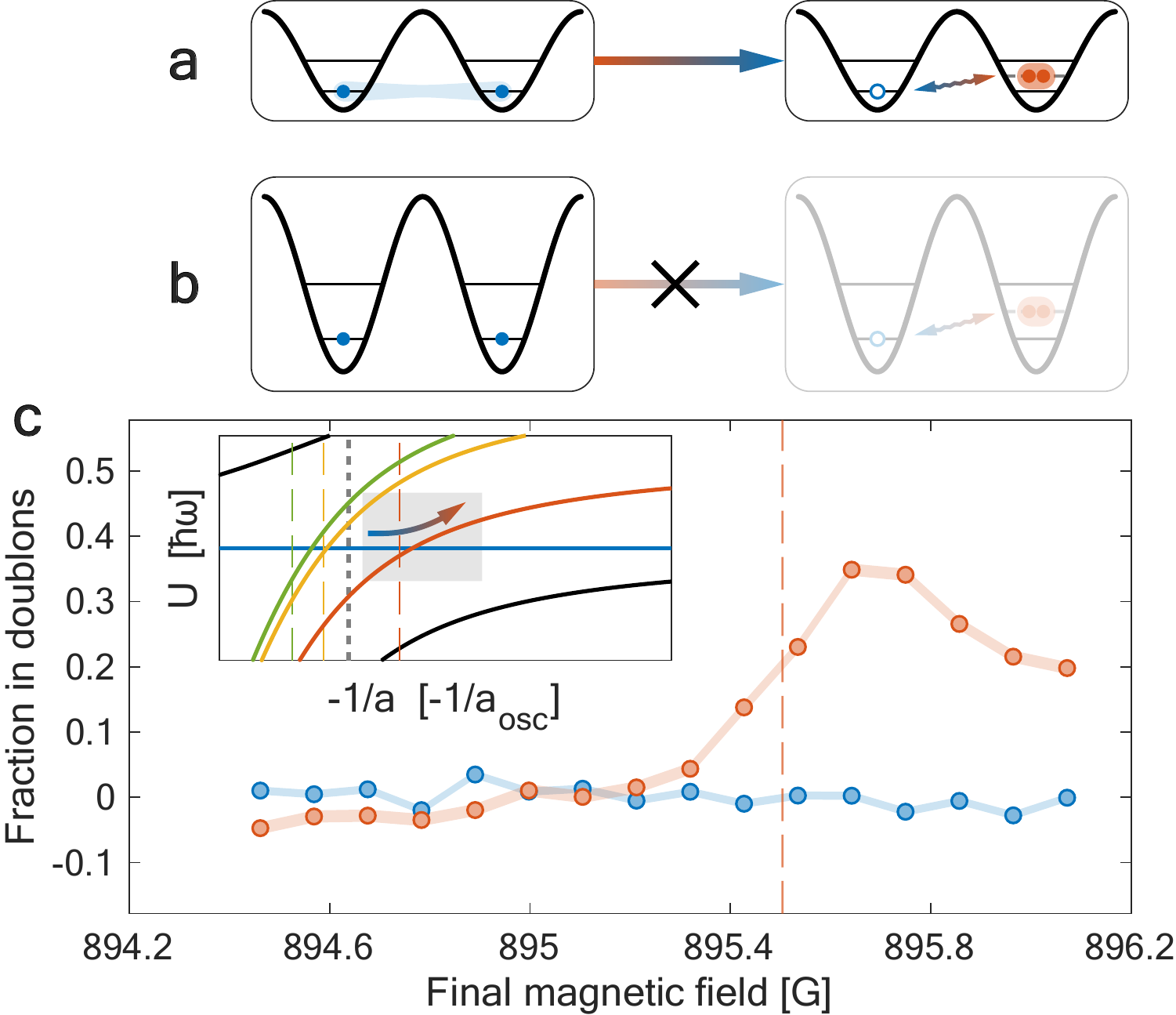}
    \caption{{\bf Creation of doublons}. 
    Starting from a singly-occupied Mott insulator of $b$ atoms in a shallow lattice $V_z=10.7\,E_R$ ({\bf a}) or a deep lattice $V_z=35\,E_R$ ({\bf b}), we sweep the magnetic field semi-adiabatically in $15\,{\rm ms}$ from $894.2\,{\rm G}$; the final field is varied while the sweep time is kept constant. We image the created ${bb}$ doublons after hiding the singlon ${b}$-atoms in the ${a}$-state by an RF pulse. We normalize to the total atom number  ({\bf c}). Doublons are created when the the field is swept past the new resonance at  $895.504\,{\rm G}$ (vertical red dashed line) when the lattice depth allows for tunnelling at $10.7\,E_R$ (red circles), while no doublons are created when tunnelling is negligible at $35\,E_R$ (blue circles). Shaded blue and red areas in {\bf c} represent the statistical errors of the mean. The reduction in doublon fraction for fields $895.8\,{\rm G}$ and higher can be explained by the finite duration of our sweep.}
    \label{fig:molecules}
\end{figure}

\noindent 
{\bf Sweeping through the new resonances}. The observed dispersive features (Figs.~\ref{fig:DeltaVsField} and \ref{fig:Jbb_vs_Ubb}) are an indirect observation of virtual molecules. In the following we directly observe real molecules by converting a Mott insulator to excited-band molecules by using an adiabatic sweep. Starting in a Mott insulator of one ${b}$ atom per site, we performed a magnetic field sweep from below to above the new resonance at $895.505\,{\rm G}$, without crossing the bulk resonance.  Such sweeps created a high fraction of doublons (Fig.~\ref{fig:molecules}). This is in contrast to sweep experiments in the bulk, where such a sweep leads to molecules which are unstable against dissociation. This observation confirms the degeneracy between singlons on different sites to a doublon on one site, coupled via a tunnelling process. A reverse sweep from above to below the resonance also created a large fraction ($\sim0.3$) of doublons. We repeated these sweep experiments for a Mott insulator of $a$ atoms and found an analogous high pair conversion for sweeps around $739.1\,{\rm G}$ without crossing the $aa$ Feshbach resonance at $737.9\,{\rm G}$. Thus, this resonance is general to any atomic species with strong interactions. 

We expect such sweeps to create highly-correlated many-body states consisting of only holes and doublons and no singly occupied sites. Our experiment bears a resemblance to sweeps of magnetic field gradients across the value $U/\lambda_{\rm lat}$, which was shown map to the Ising model in a transverse field \cite{Sachdev2002,Simon2011}. In these works, the sweep drives a quantum phase transition to antiferromagnetic order. However, without the directionality of the gradient applies, we do not expect to see the same antiferromagnetic order. These doublons are in an excited Bloch band and may have short lifetimes.  However, the lifetime may be long enough to explore interesting new phases of matter as experimentally shown in \cite{Muller2007,Vargas2021}.\\

\noindent {\bf Discussion}.
The physics of atoms in optical lattices near a Feshbach resonance is typically messy because multi-band treatments are often required. However, we demonstrate that we can perform strongly interacting physics near Feshbach resonances at scattering lengths as large as $15,000\,a_0$ {\it without} having to resort to complicated models because we use an $N\,{=}\,1$ Mott insulator where doubly occupied sites occur only virtually. Using techniques from previous work \cite{Jepsen2022}, we probe spin dynamics very close to a Feshbach resonance to study strongly interacting systems in optical lattices and observe two new contributions to the dynamics: off-site interactions, which have been predicted but were thought too small to be observed, and superexchange via virtual molecules, which have not been considered before. 

The new Feshbach resonances are unique in the sense that they involve coupling to excited bands in the center-of-mass motion.  For two atoms trapped in a harmonic potential, interactions couple only to excited bands of the relative motion due to the separability of COM and relative motion. Transfer of atoms to excited harmonic oscillator states of the relative motion via sweeps across a Feshbach resonance has been demonstrated for a fermionic band insulator \cite{Kohl2006}.  In principle, anharmonic corrections would mix relative and COM motion, and RF spectroscopy starting with two atoms on the same  site could reveal some of the energy structure in Fig.~\ref{fig:trap_branches}b.  However, due to parity selection rules, anharmonicities alone would not allow couplings to an {\it odd} parity COM state which is responsible for the strongest feature observed in our experiment. Coupling to the odd COM state is only possible due to the asymmetry of the  tunnelling process (where particles tunnel either from the left or the right side).  Similar anticrossings have been theoretically explored for double- \cite{Jachymski2020} and triple-well \cite{Schneider2009} potentials. The nature of the parity-symmetry breaking can be already seen in the limit of small interactions $|a|/a_{\rm osc} \ll 1$ where we can still use non-interacting Wannier functions to describe the wavefunctions in leading order. The lobe of the Wannier function on its neighbouring site is asymmetric, and therefore coupling to odd and even excited bands of the COM motion is non-vanishing. RF spectroscopy for two atoms per site has so far not revealed any signature for excitations of COM bands \cite{Kohl2006,Ospelkaus2006,Amato-Grill2019,Secker2020}, possibly because for the lattice depths used the anharmonicities (and therefore couplings between COM and relative motion) are too small. In principle, there should be an infinite number of Feshbach resonances involving even higher COM bands.  We have observed the three lowest bands and expect that higher resonances will be weaker as they involve more tightly bound molecular states which have only a small overlap matrix element with two unbound atoms.

Near the new resonances, the Mott insulator and the spin model will break down since they allow resonant tunnelling and therefore density fluctuations. It is an intriguing question if this could be described as a quantum phase transition into a superfluid state of doublons. With adiabatic sweeps, we verified that these new resonances can induce a transition from a Mott insulator to a state with enhanced population of doublons. \\

\noindent {\bf Acknowledgements}.
We thank Paul Niklas Jepsen for experimental assistance, discussions, and comments on the manuscript, Jinggang Xiang for comments on the manuscript, and Martin Zwierlein for sharing equipment. We acknowledge support from the NSF through the Center for Ultracold Atoms and Grant No. 1506369,  the Vannevar-Bush Faculty Fellowship, and DARPA. Y.~K.~L.~is supported in part by the National Science Foundation Graduate Research Fellowship under Grant No.~1745302.  Some of the analysis was performed by W.K. at the Aspen Center for Physics, which is supported by NSF grant PHY-1607611.
{\bf Author contributions:}  Y.~K.~L., H.~L., and W.~K. conceived the experiment. Y.~K.~L. and H.~L. performed the experiment. Y.~K.~L. and H.~L. analyzed the data. All authors discussed the results and contributed to the writing of the manuscript.
{\bf Competing interests:} The authors declare no competing financial interests. 
{\bf Data and materials availability:} The data that support the findings of this study are available from the corresponding author upon reasonable request.

\bibliographystyle{ieeetr}
\bibliography{References.bib}

\newpage

\clearpage

\setcounter{figure}{0}
\makeatletter 
\renewcommand{\thefigure}{S\@arabic\c@figure}
\makeatother

\onecolumngrid
\section*{Supplementary material}

\section{Calculating Hubbard parameters and higher-order band corrections}
The parameters in Eq.~\eqref{eq:heisenberg} are calculated using integrals \cite{Luhmann2012,Jepsen2020}
\begin{equation}
    \begin{aligned}
      U_{\sigma\sigma'} 
        &= g_{\sigma\sigma'} \int d^3 \vec{r} \ w^*(\vec{r}) w^*(\vec{r}) w(\vec{r}) w(\vec{r}) = g_{\sigma\sigma'} I_1\\
      V_{\sigma\sigma'} 
        &= g_{\sigma\sigma'} \int d^3 \vec{r} \  w^*(\vec{r}) w^*(\vec{r} - \delta\vec{r})w(\vec{r})w(\vec{r}-\delta\vec{r}) = g_{\sigma\sigma'} I_2 = \kappa U_{\sigma\sigma'}\\
      t_{\sigma\sigma'} 
        &= t + s_{\sigma\sigma'} \\
      s_{\sigma\sigma'} 
        &= -g_{\sigma\sigma'} \int d^3 \vec{r} \ w^*(\vec{r} - \delta \vec{r}) w^*(\vec{r}) w(\vec{r}) w(\vec{r}) = g_{\sigma\sigma'} I_3 = \alpha U_{\sigma\sigma'} \\
      t &= - \int dz \ w^*_z(z) \left[ \frac{\hbar^2}{2m} + V_0 \sin^2(kz) \right] w_z(z)
    \end{aligned}
    \label{eq:BHparameters}
\end{equation}
where $g_{\sigma\sigma'}$ is the coupling strength. We parameterize the magnitude of the overlap integrals of $s_{\sigma\sigma'}$ and $V_{\sigma\sigma'}$ with respect to $U_{\sigma\sigma'}$ with constants $\alpha$ and $\kappa$
\begin{equation}
    \begin{aligned}
    \alpha & = -I_3 / I_1 & \approx 2.868\times10^{-3}\\
    \kappa & = I_2 / I_1 & \approx 6.915\times 10^{-5}
    \end{aligned}
    \label{eq:franckCondonFactors}
\end{equation}
The negative sign in the expression for $\alpha$ is added to make it a positive quantity since $I_3<0$. The interaction strength $g_{\sigma\sigma'}$ determines the strength of each on-site $U$, off-site $V$, and bond-charge tunnelling $s$. Equations \eqref{eq:BHparameters} have been derived in the weak coupling limit where  the coupling strength is proportional to the scattering length ($g_{\sigma\sigma'} \propto a_{\sigma\sigma'}$).  For strong interactions, the interaction of two particles per site becomes unitarity limited.  For an isotropic lattice, this implies that $a_{bb}$ effectively saturates at $a_\mathrm{osc}$.  

There is no existing theory that describes the behavior of bond-charge tunnelling $s$ and off-site interactions $V$ for strongly interacting pairs \cite{Schneider2009}. In this paper, we propose a way to phenomenologically generalize eq.~\eqref{eq:BHparameters} for strong couplings by noting that $s$, $U$, and $V$ all arise due to $\delta$-function interactions and finite wavefunction overlap; in the case of $s$ and $V$, the interaction is approximately determined by the \enquote{leakage} of a Wannier function onto its neighbouring site. Thus, we suggest that the effective interaction parameter  $g_{\sigma\sigma'}$ can be used to parameterize $s$, $U$, and $V$ by assuming that the ratios between $I_1, I_2,$ and $I_3$ are independent of interactions.  

Using this idea, we may write everything in terms of $U_{bb}$ with appropriate scaling factors $\alpha$ and $\kappa$ from \eqref{eq:franckCondonFactors}:
\begin{equation}
    \begin{aligned}
    -J_{bb} 
    &= \frac{-4t_{bb}^2}{U_{bb}} \\
    &= \frac{-4(t + s_{bb})^2}{U_{bb}} \\
    &= \frac{-4t^2 - 8 t (\alpha U_{bb}) - 4(\alpha U_{bb})^2}{U_{bb}} \\
    &= -\frac{4t^2}{U_{bb}} - 8\alpha t - 4\alpha^2 U_{bb} 
    \end{aligned}
\end{equation}
For a harmonic approximation, the above analysis holds for bands regardless of their COM excitation. The off-site interaction of the lowest COM state is
\begin{equation}
    2 V_{bb} = 2 \kappa\, U_{bb}
\end{equation}
Mapping onto Eq.~\eqref{eq:Jbb}
\begin{equation}
\begin{aligned}
    J_{bb}^* 
    & = -J_{bb} + 2V_{bb} \\
    & = -\frac{4t^2}{U_{bb}}-8\alpha t - 4\alpha^2 U_{bb} + 2 \kappa U_{bb} \\
    & \approx -\frac{4t^2}{U_{bb}} + (2 \kappa - 4 \alpha^2) U_{bb} 
\end{aligned}
\label{eq:JbbAnalysis}
\end{equation}
where we have omitted the second term in the second line of \eqref{eq:JbbAnalysis} since it is a small constant we can subtract. The coefficient of the last term is, for lattice depths of $V_{x,y,z}=35\,E_R,\,35\,E_R,\,10.7\,E_R$:
\begin{equation}
    2\kappa - 4\alpha^2 \approx 1.524 \,\kappa
\end{equation}
This implies that the off-site interactions dominate over the bond-charge contribution by a factor of four. So we expect the off-site term in $J_{bb}^*$ to be approximately $B\cdot\kappa\,U_{bb} \approx 1.5 \cdot \kappa\,U_{bb}$. Our fitted value of $B=1.3(4)$ is in good agreement with this prediction.

When we subtract contributions that depend on the ${aa}$ and ${ab}$ scattering lengths we modify $U_{aa}$ and $U_{ab}$ to include the coupling to higher-order bands. These are second-order shifts to the energies due to off-resonant couplings to these higher bands. See \cite{Campbell2006,Luhmann2012,Jepsen2020,Jepsen2022} for more details.

\end{document}